# Small scale magnetosphere: Laboratory experiment, physical model and Hall MHD simulation

**I F Shaikhislamov, V M Antonov, Yu P Zakharov, E L Boyarintsev, A V Melekhov, V G Posukh and A G Ponomarenko**

Dep. of Laser Plasma, Institute of Laser Physics SB RAS, pr.Lavrentyeva 13/3, 630090, Novosibirsk

E-mail: ildars@ngs.ru

**Abstract**
A problem of magnetosphere formation on ion inertia scale around weakly magnetized bodies is investigated by means of laboratory experiment, analytical analysis and 2.5D Hall MHD simulation. Experimental evidence of specific magnetic field generated by the Hall term is presented. Direct comparison of regimes with small and large ion inertia length revealed striking differences in measured magnetopause position and plasma stand off distance. Analytical model is presented, which explains such basic features of mini-magnetosphere observed in previous kinetic simulations as disappearance of bow shock and plasma stopping at Stoermer particle limit instead of pressure balance distance. Numerical simulation is found to be in a good agreement with experiments and analytical model. It gives detailed spatial structure of Hall field and reveals that while ions penetrate deep inside mini-magnetosphere electrons overflow around it along magnetopause boundary.
**PACS:** 94.30.C+52.30.Cv

## 1. Introduction

In recent years there emerged a number of related problems dealing with a mini-magnetosphere. Mini-magnetosphere forms when a small body like asteroid, or localized surface region like on Moon or Mars, or a spacecraft possesses an intrinsic magnetic field. The term mini signifies a specific spatial scale at which it applies – ion gyro-radius or ion inertia length $c/\omega_{pi}$. At these scales interaction of Solar Wind with a localized magnetic field is different from the well-known planetary magnetospheres because of two-fluid and kinetic effects.

A general problem of plasma-field interaction at ion scales has a long history and a number of breakthroughs could be attributed to its application. AMPTE barium releases in the Earth magnetosphere (Bernhardt *et al* 1987) and related laboratory experiments (Okada *et al* 1981, Zakharov *et al* 1986) revealed a new kind of Raleigh-Taylor instability of plasma boundary (Hassam and Huba 1987) driven by the so called Hall term $\mathbf{J} \times \mathbf{B}/nec$. In magnetic reconnection research a long standing problem of diffusion region localization has been resolved after taking into account the Hall term (Mandt *et al* 1994). In technology Hall thrusters (Mikhailichenko *et al* 1973) and plasma switches (Fruchman and Maron 1991) utilize a two-fluid regime in which electrons flow distinctly apart from ions.

Galileo spacecraft encounter with asteroid Gaspra in 1991 and asteroid Ida in 1993 motivated studies of specific signatures that a weakly magnetized body produces in SW. It was recognized that under ion scales incompressible whistler modes would dominate instead of magnetosonic waves. Though peculiar magnetic signals of Galileo were eventually interpreted as SW discontinuities (Blanco-Cano *et al* 2003) because were registered too far away from the asteroids (>1000 km), performed Hall MHD and hybrid simulations (Omidi *et al* 2002) revealed how distinctly a mini-magnetosphere is different. Whistler and magnetosonic wake is generated behind the body while ahead there is no ion deflection and density pile up. A shocked upstream region and a strong obstacle to SW roughly resembling magnetosphere appear only when pressure balance stand off distance becomes larger than the ion inertia scale.

Since the discovery of lunar crustal magnetic fields in Apollo missions, their mapping by Lunar Prospector gave ample examples that SW does interact with lunar magnetic anomalous. A number of spacecrafts have observed at altitudes as high as 100 km magnetic enhancements, particle fluxes and waves resembling a bow-shock structure undoubtedly associated with crustal magnetic sources. On Moon a mini-magnetosphere might be useful as a shield against SW plasma and unusual albedo markings have been found around several lunar magnetic anomalies. However, extensive search out of more than a thousand flybys over Crisium Antipode anomaly revealed only two cases of actual density drops that could be expected inside of magnetosphere (Halekas *et al* 2008). Characteristically, unique solar wind conditions with unusually small ion inertia length (~57 over average 97 km) were associated with them. Preliminary results of SELENE Explorer (Saito *et al* 2010) revealed distinct magnetic reflection of SW ions over South Pole Aitken anomalies accompanied sometimes with heated electrons and slight deceleration of incoming ions. Correlated absence of surface deflected ions indicated existence of shielded regions.

Besides mentioned examples, Phobos-2 mission in 1989 gave evidence that Phobos might be magnetized generating a draping of magnetic field at distances 200-300 km depending on SW density (Mordovskaya *et al* 2001). It was argued that observed magnetic variations could be in fact due to Mars exosphere, so future missions should clarify this. In (Shabansky *et al* 1989) it was proposed to employ a super-conductive magnet on board of a satellite to perform various geophysical experiments. This will soon be realized on International Space Station after installation of AMS magnet. It will interact with rarified ionosphere in kinetic regime (Zakharov *et al* 2009) as the stand off distance estimates about ~10 m, gyro-radius ~40 m and inertia length ~1 km.

The idea to shield a spacecraft from energetic cosmic radiation by on-board magnetic field source was put forward as early as in 60-s. Technically feasible source (for example, Spillantini *et al* 2000) with effective moment $\sim 10^9 \div 10^{10} \text{A} \cdot \text{m}^2$ would create in SW a mini-magnetosphere ~1 km size much smaller than ion scales ~100 km. Recently, a dipole magnetic field as a shield was tested in laboratory in kinetic regime (Bamford *et al* 2008). However, experimental conditions of sub-Alfvenic and geometrically very narrow plasma beam guided by strong external magnetic field toward the dipole might be relevant only to ionosphere but not SW. In hybrid numerical simulation of this experiment (Gargate *et al* 2008) performed under more realistic super-Alfvenic conditions it is reported, contrary to many other simulations, that ions are deflected at the expected stand off distance and density cavity of the corresponding size is formed, even while ion gyro-radius and inertia length are order of magnitude larger than the mini-magnetosphere size.

Perhaps most revealing and fundamental results, at least for the topic of present work, came from numerous numerical studies of rather controversial idea of magnetic sail proposed by (Winglee *et al* 2000). Parametric study by hybrid simulation (Fujita 2004) showed that the size of mini-magnetosphere is equal to MHD stand off distance when ion inertia length is small and to a Stoermer radius otherwise with a sharp transition in between. Thus, in kinetic regime plasma behaves like individual orbiting particles. How it happens at conditions of strongly collective plasma wasn't analyzed in any of the works.

Laboratory modeling is another useful and independent way to study the physics of mini-magnetosphere. In the paper we present results of several terrella experiments which cover sufficiently large range of kinetic scales. In the first one plasma flow consisting of hydrogen ions was used and the ion scales were about twice smaller than the stand off distance. A well defined magnetosphere with plasma cavity was observed. Systematic measurements in meridian plane revealed for the first time existence of global out of plane component of magnetic field. The bipolar structure of this field with two opposite maxima in meridian plane is totally different from what could be generated by convection term $\mathbf{V} \times \mathbf{B}$ and indicates its origin due to the Hall term $\mathbf{J} \times \mathbf{B}/nec$. In other experiment employing instead of hydrogen much heavier Argon ions made it possible to achieve ion scales several times larger than the stand off distance. While magnetic barrier and magnetopause current were still observed, though quite farther than the stand off distance, plasma penetrated inside magnetosphere all the way down to dipole cover. It should be noted that a number of earlier terrella experiments were carried out in kinetic regime (for example, LG Cohen and SKF Karlsson 1969). However, data that could shed light on the problem under consideration are unavailable, probably because the aim of earlier experiments was to model Earth-like magnetosphere.

Presented experimental results as well as sited above numerical simulations impose fundamental questions why plasma penetrates through magnetopause and how it moves inside magnetosphere without advecting dipole field. We propose that magnetic field generated via Hall term and related Hall current are behind the unusual properties of mini-magnetosphere. Hall current is maximal at the axis of interaction and is directed along SW velocity. As the dipole magnetic field is advected by a combination $(\mathbf{V} - \mathbf{J}/ne) \times \mathbf{B}$ (namely by electrons) ions have to penetrate magnetic barrier to cancel the Hall current. To demonstrate the idea we developed simple analytical model which helps to estimate the value of Hall field, penetration velocity and magnetopause position in dependence on ion inertia length. For a quantitative characterization we employ 2.5-dimension Hall MHD numerical simulation that includes Hall term on the one hand and bow shock physics on the other. When Hall term is switched on, out of plane magnetic field is generated, the bow shock disappears, while plasma penetrates into the mini-magnetosphere and is eventually stopped at Stoermer radius. It was found that when ion scales are much larger than system size plasma tends to cancel all current $\mathbf{V} \approx \mathbf{J}/ne$. Thus, plasma moves in strong magnetic field inside mini-magnetosphere like orbiting particles. The other novel feature is that electrons flow essentially differently from ions. They don't penetrate magnetosphere and overflow it along magnetopause boundary.

The aim of the work is to build a comprehensive picture of mini-magnetosphere based on laboratory experiments, analytical model and numerical study. We mostly discuss a frontal part of magnetosphere at condition of absent Interplanetary Magnetic Field as dictated by experimental data and by necessary simplifications of analytical model. However, with the aim of comparing with results of PIC kinetic codes, in numerical simulations we also study the far tail and far upstream regions in presence of oblique IMF. The paper consists of five sections. In the second section four terrella and one laser-produced plasma experiment are described. Next a model demonstrating the physics involved is presented. In the forth section results of 2.5D Hall MHD simulation are described, followed by discussion and conclusions.

## 2. Experiments

Throughout the paper GSM coordinate system is used. In the first experimental set up theta-pinch plasma interacts with magnetic dipole of moment $\mu = 1.25 \cdot 10^5 \, \mathrm{G \cdot cm^3}$. Stainless dipole cover has the radius of $3.75 \, \mathrm{cm}$. Operating time of theta-pinch and dipole is $100 \, \mu s$ and $0.5 \, s$ respectively. After a time of about $20-50 \, \mu s$ following discharge steady state magnetosphere with spatial scale $\approx 10 \, \mathrm{cm}$ is formed. Large range of the kinetic scales in relation to Terrella size was achieved in physically different ways: varying plasma density, velocity and employing light Hydrogen and heavy Argon ions. Following (Omidi et al 2002) we define the Hall parameter as a relation of the pressure balance stand off distance $R_M = \left(\mu^2 / 2\pi n_i M V_o^2\right)^{1/6}$ to the ion inertia length $L_{pi} = c/\omega_{pi}$, $D = R_M/L_{pi}$. In the table specific

| № | $n_i$, cm$^{-3}$ | $V_o$, km/s | Ion | D |
|---|---|---|---|---|
| 1 | $4 \cdot 10^{13}$ | 40 | H$^+$ | 3.3 |
| 2 | $2 \cdot 10^{13}$ | 100 | H$^+$ | 1.9 |
| 3 | $1.5 \cdot 10^{12}$ | 120 | H$^+$ | 0.75 |
| 4 | $2 \cdot 10^{12}$ | 50 | Ar$^{4+}$ | 0.4 |

conditions of four regimes are presented. For all of them the flow is super-sonic ($M_s \approx 3$) and super-Alfvenic ($M_A \approx 3-7$, weak background magnetic field is applied along X axis to direct plasma from theta-pinch to dipole). Magnetic Reynolds number is $\geq 10$, Knudsen number $\geq 5$.

In the regime №2 detailed measurements of magnetic field and plasma density were performed by mapping meridian, equator and terminator planes. Regions with size $-25 \leq x \leq 25$, $-8 \leq z \leq 25$, $-25 \leq y \leq 25 \, \mathrm{cm}$ were covered more or less uniformly by about 450 points in each plane. The measurement grid was dense enough to draw the structure of magnetosphere. In fig 1-A one can see meridian plane where magnetic field lines are mapped over density plot of Chapman-Ferraro current $J_y$. In the fig 1-B plot of plasma density is presented. Only experimental data and smoothing and interpolating procedures were used to draw the pictures. One can see the essential features of magnetosphere – Chapman-Ferraro current, cusps, tail, density cavity. It should be noted that though there is density increase near the stagnation point, no bow shock develops in the experiment. Detailed mapping revealed the existence of out of plane $B_y$ component of magnetic field (figure 1-C) that could not be explained in the MHD frame. It is positive in the North hemisphere. In the South part measurements were made only down to $z = -8\,\text{cm}$. It was enough to see that $B_y$ is negative and asymmetric in respect to Z axis: $B_y(z) = -B_y(-z)$. Its maximum value $\approx 50\,\text{G}$ is about 4 times smaller than the jump of field at the magnetopause $\Delta B_z \approx 200\,\text{G}$. The structure of $B_y$ field closely follows the Chapman-Ferraro current.

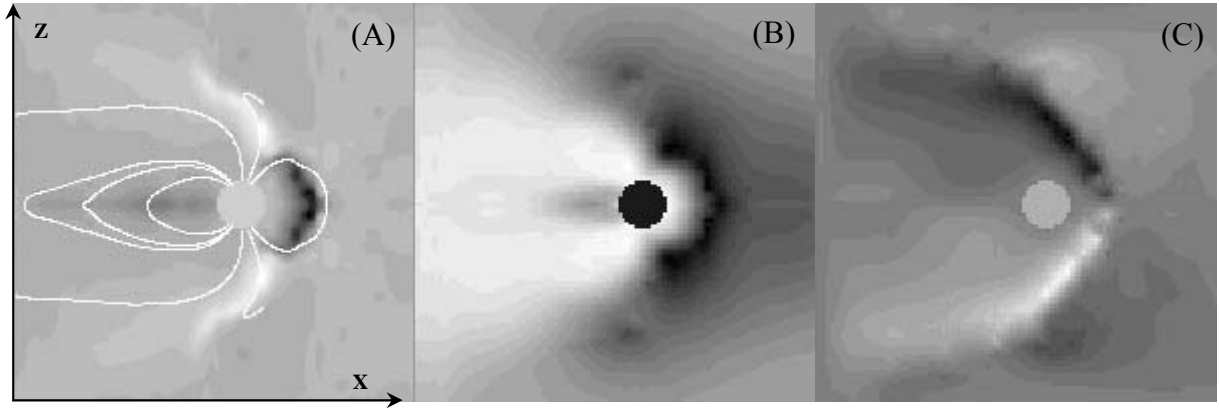

**Figure 1** Meridian structure of laboratory magnetosphere. Spatial size of region is 70×70 cm. Circle marks the dipole cover.
**A** - Density plot of current $J_Y$. Maximum black (white) corresponds to 96 A/cm$^2$ (- 40 A/cm$^2$). White lines show magnetic field lines.
**B** - Plot of plasma density $n_I$. Maximum black is $3.4 \cdot 10^{13}\,\text{cm}^{-3}$, white - zero.
**C** – Density plot of magnetic field component $B_Y$. Maximum black (white) is ±54 G.

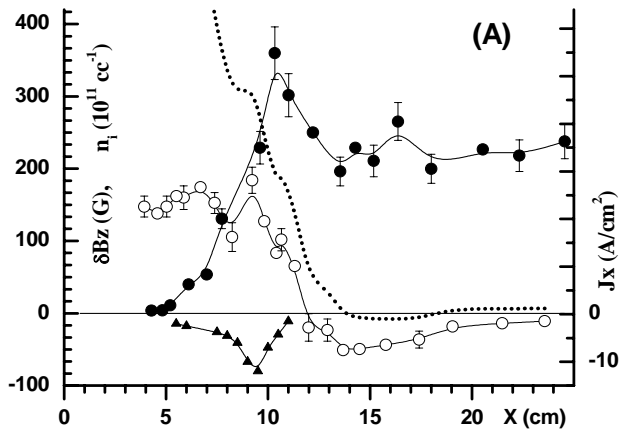

**Figure 2** Profiles of total magnetic field (dotted curve), magnetic field perturbation (o), plasma density (●) and electric current $J_x$ (▲) along X axis measured in the regime № 2.

In figure 2 profiles of density and main field component along the plasma flow direction are shown. It could be seen that boundary layer is about 3÷4 cm wide (from maximum to minimum of field perturbation) and magnetopause, defined as maximum of Chapman-Ferraro current, is positioned at $x \approx 11.5\,\text{cm}$. Dipole field is practically absent upstream of magnetopause. Density front and current $J_x \sim \partial B_y / \partial z$ associated with out of plane $B_y$ field penetrate beyond magnetopause by a few cm. The value of current velocity $J_x / ne = -(20 \div 40)\,\text{km/s}$ is about three times smaller than upstream plasma velocity $V_0$.

In the regime listed as №1 Hall parameter was the largest due to greater density and lower velocity, while in №3 it was smaller than unity for the reverse reasons. In the forth experiment argon instead of hydrogen was used. Because of large atomic mass it yielded the smallest Hall parameter which for the ion charge $Z = 2 \div 4$ is in the range $D = 0.2 \div 0.4$. In the figure 3 profiles of magnetosphere along x axis are shown for all regimes. Solid vertical line in each panel indicates a "sub-solar" stand off distance $R_M$ calculated by theoretical formula. Dashed lines indicate measured magnetopause position $R_m$ and boundary of plasma penetration inside magnetosphere $R_p$. For the largest D (upper panel) theoretical and measured magnetopause positions are very close to each other and plasma doesn't penetrate beyond the boundary layer. The field jump at magnetopause $\Delta B_z \approx 150\,G$ is capable to balance flow with velocity of $\Delta V = \sqrt{\Delta B_z^2/(8\pi n_i M)} \approx 37\,km/s$ which is practically equal to the upstream value $V_0$. On the other hand the regimes with $D<1$ (two bottom panels) exhibit different structure. Magnetopause is significantly farther from the dipole than expected (by a factor of 1.5 for Argon) and plasma penetrates deep inside magnetosphere. The field jump at the magnetopause is small $\Delta B_z \leq 50\,G$ and is capable to balance flow with velocity of only $\approx 40\,km/s$ for regime №3 and $\approx 10\,km/s$ for №4. This is several times smaller than $V_0$ for both cases. Thus, plasma should penetrate through magnetic barrier with little deceleration and it does so as could be judged from the density profiles. For Argon experiment plasma hits the dipole cover at $x = 3.75\,cm$, so there is no cavity at all. In test particle model the closest ion approach at the X axis is equal to 0.6 of Stoermer radius and calculates as 6.7, 6.1 and 3 cm for regimes №2, 3 and 4 respectively. These values are consistent with observed $R_p$. The second regime is clearly intermediate between large and small Hall parameters. While it shows plasma penetration beyond the boundary layer, the magnetic barrier is strong enough to stop the flow with velocity $\Delta V \approx 70$ which is comparable to $V_0 = 100\,km/s$.

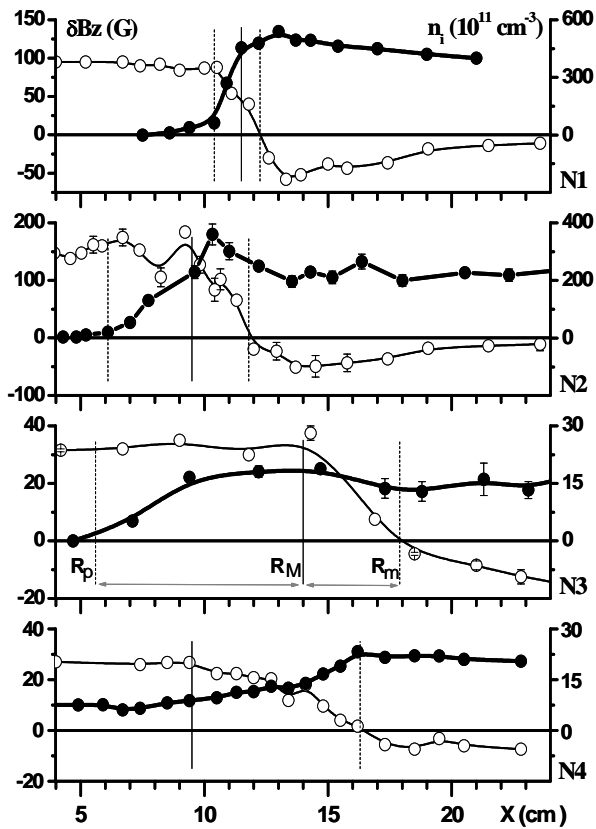

**Figure 3** Profiles of magnetic field perturbation (o) and plasma density (●) measured for various regimes. Thin vertical line indicates a "sub-solar" stand off distance calculated by theoretical formula. Dashed lines indicate measured magnetopause position and boundary of plasma penetration inside magnetosphere.

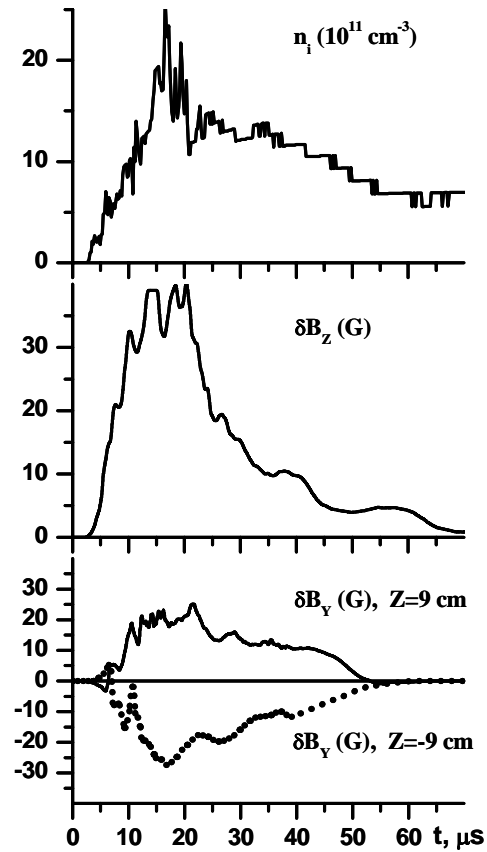

**Figure 4** Oscilloscope signals inside magnetosphere at x=9 cm in the regime №3.
Upper panels - density and variation of magnetic field at z=0.
Lower panel – $B_Y$ component measured above and below equator.

In the regime №3 out of plane magnetic field $B_y$ has been found to exist well inside magnetosphere approximately in the whole region where penetrated plasma was observed. In figure 4 oscilloscope signals of density and main field component $\delta B_z$ measured at x=9 cm are shown. From figure 3 (№3) one can judge that this position is far downstream of magnetopause and close to the boundary of plasma cavity. There is a quasi-stationary phase of interaction between 10 to 20 μs when magnetic field inside magnetosphere produced by compression is in the range $\delta B_z = 30 \div 40\,G$. In the bottom panel $B_y$ signals are shown. One above equator is positive while below it is negative, like in the figure 1-C. Dynamically $B_y$ component more or less follows $\delta B_z$ signal. Estimation of current velocity gives $J_x/ne \approx -75$ which is smaller but comparable to the upstream plasma speed $V_0 = 120\,km/s$.

In the second experimental set up laser-produced plasma instead of theta-pinch is used. Details of the laser plasma experiments can be found in (Ponomarenko *et al* 2008). Two laser beams are focused on a solid target placed at a distance of 66 cm from the dipole center. Produced plasma consists of hydrogen and carbon ions in approximately equal parts with estimated average ion charge $\langle Z \rangle \approx 2$ and average ion mass $\langle M \rangle \approx 5.5$. It expands in a cone ~1 sr with velocity about $V_0 \approx 100\,km/s$. In the interaction region measured density is $n_i \approx 5 \cdot 10^{11}\,cm^{-3}$ and the stand off distance estimates as $R_M \approx 16\,cm$, while ion inertia length as $L_{pi} \approx 40\,cm$. Hall parameter is sufficiently small $D \approx 0.4$. In the upper panel of figure 5 ion current measured by probe is shown. Flow consists of two pressure jumps which is a consequence of specific pulse and tail mode of laser amplifier. We will study the interaction preceding the second pressure jump. The time of interest is marked by dashed vertical line.

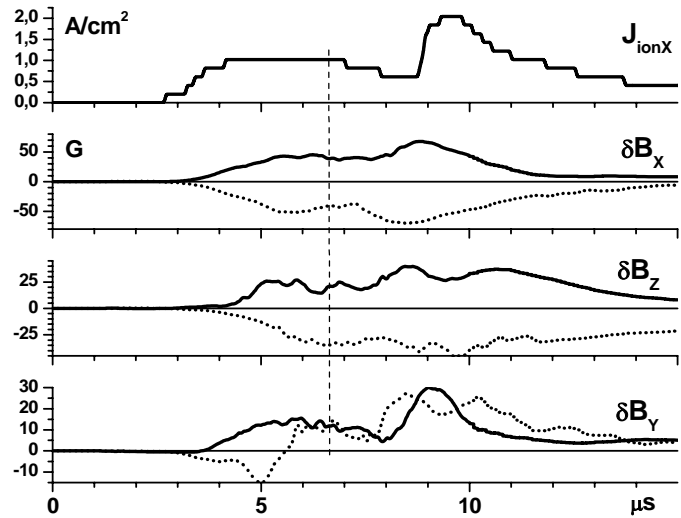

**Figure 5** Upper panel - ion current measured at X=12 cm in the absence of dipole field.
Next panels show three components of magnetic perturbation at position above equator Z=4 cm. Solid curves – for normal Southward direction of magnetic moment, dotted – at reversed Northward direction.

At the next panels of figure 5 there are shown three components of magnetic field perturbation measured in meridian plane Y=0 above equator at position X=12 cm, Z=4 cm. One can see that steady plasma flow produces more or less stationary magnetosphere. Positive values of $\delta B_x$ and $\delta B_z$ signals (solid curves) correspond to flattening and compression of dipole field at dayside sector. One can see also that there is out of plane component $B_y$. The major finding of this experiment is $B_y$ behavior at reversing the dipole moment. Signals in case when magnetic moment is changed to opposite polarity (North direction) are shown by dotted curves. As expected, perturbations of dipole field $\delta B_x$ and $\delta B_z$ change sign. $B_y$ component also exhibits brief initial reversal. However, during most of the interaction time it shows the same polarity. This is a persistent feature checked in several shots. Thus, out of plane field is of quadratic and non-MHD nature. It could be deduced also that characteristic generation time of non-MHD process is about 2 μs, which is close to a typical time of flight $R_M/V_o$.

Spatial profile of magnetic field perturbation along the interaction axis is presented in the figure 6. It is similar to what has been observed with theta-pinch plasma. Magnetopause position $R_m \approx 20$ cm is by 4 cm ahead of estimated pressure balance distance. At position X=12 cm, which is well inside magnetosphere, magnetic probe measured distribution of the out of plane component $B_y$ along Z axis. This is shown in figure 7. One can see that it definitely changes sign at equator crossing. The linear fit (dashed line) gives estimation for electric current density $J_x \approx 1\,A/cm^2$, which is equal to ion current measured by probe (figure 5).

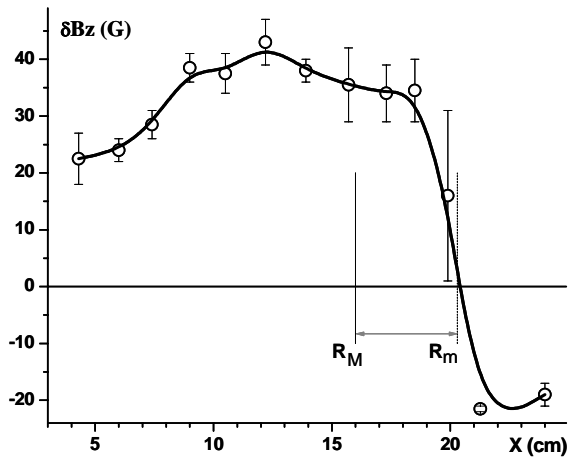
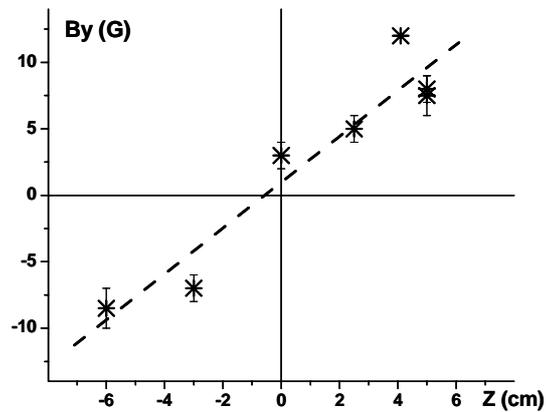

**Figure 6** Profile of magnetosphere measured in laser-produced plasma experiment.

**Figure 7** Out of plane magnetic field distribution along Z axis measured inside magnetosphere at X=12 cm.

### 3. Preliminary analysis

In MHD frame interaction of plasma flow with magnetic dipole is characterized by generation of Chapman-Ferraro current which decelerates plasma at a pressure balance distance and forms a magnetosphere. Magnetosphere contains the dipole field and external plasma doesn't penetrate inside it beyond thin boundary layer. However, presented experiments revealed that in kinetic regime plasma penetrates deep inside the region of dipole field. The same picture has been shown explicitly in a number of numerical simulations (for example, Blanco-Cano *et al* 2004). This raises a fundamental question how plasma moves across field. It can't be answered by invoking anomalous resistivity due to micro-instabilities because no definite signs are observed to that effect. One can estimate that fast diffusion requires anomalous collisions with exceptionally high rate of the order of electron gyrofrequency. Individual ions may penetrate across magnetic boundary by a distance of gyroradius, but it doesn't explain why magnetic field isn't advected by a mean plasma velocity. The logical answer is that inside magnetosphere only ions can move across field while electrons don't move.

The other fundamental issue concerns the general structure of magnetosphere. Based on numerical results it was deduced in (Blanco-Cano *et al* 2004, Fujita 2004) that no magnetosphere at all is formed at large ion gyroradius. However, a magnetopause as a boundary of dipole field should necessarily exist as presented above experiments show. Indeed, sufficiently far from the dipole its magnetic field should be totally expelled by SW, while sufficiently close it should dominate. At the boundary that divides these regions a current should flow which decelerates incoming ions. In MHD case magnetopause position is a stand off distance at which ions are totally stopped. Otherwise this position should be farther off and ions deceleration only partial. The sharpness of magnetopause current layer is regulated by electrons. While electron inertia length is sufficiently small, which is the case for most applications, the layer thickness should be also small compared to system size.

In Hall MHD the Ohm' law is generalized to include the Hall term

$$\mathbf{E} = -\frac{\mathbf{V_e} \times \mathbf{B}}{c} = -\frac{\mathbf{V} \times \mathbf{B}}{c} + \frac{\mathbf{J} \times \mathbf{B}}{nec} \tag{3.1}$$

We ignored electron pressure which is inessential for the present study. Electron velocity is expressed through ion (or plasma) velocity and current. One can see that Hall electric field is the one that decelerates plasma $E_x \sim J_y B_z$. Because of latitude dependence, Chapman-Ferraro current generates new component of magnetic field directed along it:

$$\frac{\partial}{\partial t} B_y \approx -\frac{\partial}{\partial x} V_x B_y + \frac{\partial}{\partial z}\left(V_y - \frac{J_y}{ne}\right) B_z \tag{3.2}$$

The structure of the Hall magnetic field is bipolar. Maximums are in meridian plane with positive values (in dawn to dusk direction) in the North hemisphere and negative – in the South. This is a specific feature of the Hall term that makes it distinctly different from the usual MHD. In MHD plasma flow around dipole can generate $B_y$ component of quadruple structure such that $B_y$ is zero in the meridian and equator plane. Moreover, there is another fundamental aspect of Hall field. Inverting magnetic moment leads to inversion of all magnetic fields generated by MHD processes, while Hall field doesn't change sign because of quadratic nature.

A current associated with Hall field is directed perpendicular to Chapman-Ferraro current. At the interaction axis it is maximal, $J_x \sim -\partial B_y / \partial z$, and flows like plasma towards dipole. Because of the Hall current advection of the main field component $B_z$ also changes according to

$$\frac{\partial}{\partial t} B_z \approx -\frac{\partial}{\partial x}\left(V_x - \frac{J_x}{ne}\right) B_z \tag{3.3}$$

Thus, in a stationary state plasma velocity should be equal to current velocity $V_x \approx J_x / ne$. That is, plasma should penetrate across magnetopause. We note that exactly this tendency of electric current density reaching the value of ion current has been observed in experiments at conditions of large ion inertia length. Because the jump of velocity across magnetopause decreases, kinetic pressure and jump of magnetic field also decreases. To accommodate this change magnetopause should move farther away from the dipole.

Inside the magnetosphere, current along interaction axis $J_x$ leads to the force $-J_x B_z$ that accelerates plasma in the direction of Chapman-Ferraro current. Non-zero $V_y$ velocity at the interaction axis is another specific feature of the Hall term. Taking $V_y$ component into account in (3.2) one can see that there should be current inside magnetosphere to compensate it: $J_y \approx neV_y$. Reminding the ion momentum equation $nM\, d\mathbf{V}/dt \approx \mathbf{J} \times \mathbf{B}/c$ and taking the current velocity to be equal to plasma velocity $\mathbf{J}/ne = \mathbf{V}$ we arrive to a simple conclusion that inside magnetosphere plasma should move as particles orbiting in magnetic field. Obviously, in that case the closest distance of plasma penetration in dipole field is determined by a Stoermer limit. On the other hand, at condition $\mathbf{J}/ne = \mathbf{V}$ electron velocity is zero. That is, electrons don't move inside magnetosphere.

Other fundamental aspect concerns bow shock. When supersonic flow is stopped at magnetopause a shock wave is generated upstream. At the shock plasma is heated and decelerated so as local sound speed becomes larger than flow speed. For typical supersonic SW plasma velocity drops to one forth of initial value while thermal pressure increases to $3/4\, nMV_0^2$. In the region between shock and magnetopause flow gradually drops to zero while pressure reaches kinetic pressure. However, if plasma penetrates magnetopause then the flow velocity in the magnetosheath also increases. When penetration velocity becomes faster than $V_0/4$ stationary shock cannot exist. As will be seen in simulations at these conditions bow shock upstream of magnetopause doesn't exist at all. Simple intuitive arguments given above outline the following picture. When ion inertia length is larger than pressure balance stopping distance, the single magnetosheath structure of usual MHD transforms into the double structure of a mini-magnetosphere. There is magnetopause as a boundary of dipole magnetic field where plasma velocity slightly decelerates and Hall magnetic field is generated and there is the inner boundary of the order of Stoermer radius at which plasma is eventually stopped.

Let's estimate penetration velocity $V_p$ or deceleration $\Delta V = V_o - V_p$ assuming that there is no shock ahead of the magnetopause and thermal pressure could be ignored. Suppose that magnetopause at which plasma decelerates is positioned at $R_m$. Taking into account that the jump of magnetic field at magnetopause that decelerates plasma should be twice that of the dipole field at this distance, using continuity condition $nV = const$ and ion momentum equation $Mn_o V_o \Delta V = (\Delta B_z)^2 / 8\pi$ one arrives at:

$$R_m = R_M (V_o / \Delta V)^{1/6} \tag{3.4}$$

Next, taking z-differential from (3.2) to obtain Hall current and assuming that latitude variation has the order of magnetopause radius $\partial^2 / \partial z^2 \sim -1/R_m^2$ one gets

$$\frac{4\pi}{c} \frac{\partial}{\partial x} V_x J_x \approx -\frac{J_y}{ne} B_z / R_m^2 \tag{3.5}$$

Using ion momentum equation $J_y B_z / nc = M V_x \partial V_x / \partial x$ to substitute magnetic force and integrating in respect to x we arrive to

$$2D^2 \frac{V_p J_x}{n_o e V_o^2} = \left(1 - V_p^2 / V_o^2\right) \cdot (\Delta V / V_o)^{1/3} \tag{3.6}$$

After crossing magnetopause Hall current should be equal to ion current $J_x = enV_p = en_o V_o$. Final algebraic equation for relative penetration velocity $v_p = V_p / V_o$ reads as

$$2D^2 v_p = \left(1 - v_p^2\right) \cdot (1 - v_p)^{1/3} \tag{3.7}$$

Asymptotic solution at $D \ll 1$ is $\Delta v = 1 - v_p \approx D^{3/2}$. Scaling of Hall field is given by $B_{z,max} \approx 4\pi J_x R_m / c \approx \sqrt{4\pi n_o M V_o^2} \cdot R_m / L_{pi}$.

Now we would like to develop a self-consistent numerical model which incorporates Hall effects on the one hand and such MHD features as magnetopause, bow shock and a shocked region on the other. It should be stressed that, because the kinetic scales are related through pressure balance distance, when Hall effects are strong gyroradius is correspondingly large. Thus, a fluid approach to a problem with ion gyroradius being larger than system size could not be strictly validated. However, as a reference frame Hall MHD is indispensable as a step to a more elaborated analysis. Besides, for our particular problem thermal chaotic velocity of ions is much smaller than bulk speed. As will be shown, even in the strongly kinetic limit of large gyroradius a small region around dipole where plasma is eventually deflected and thermal velocity becomes comparable to bulk speed is much smaller than characteristic system size.

For numerical simulation we reduce the problem to two dimensions with $\partial / \partial y = 0$. Note that y-components of field and velocity are not zero. While it makes it necessary to employ two-dimension line dipole with different scaling of stopping distance, such geometry is most suitable for study of Hall physics. In deriving Hall MHD electron mass is included in equations for the reason of making numerical processing more stable. The model will be made as simple as possible with the aim to reveal the physics of mini-magnetosphere, rather than to achieve exact and full description.

It should be noted that nowadays there is a number of 2D and 3D hybrid codes which might be more suitable for numerical study of the problem under consideration (see references above). Properties of mini-magnetosphere have been derived mainly from these codes. However, we believe that most important of these properties have a two-fluid nature. To understand them more clearly it is useful to employ two-fluid simulation which could be compared with much more complex hybrid simulations. The comparison will help in developing a comprehensive physical model that explains such features as plasma penetration beyond MHD stand off distance and disappearance of bow shock.

**4. Hall MHD Model**

We start from momentum fluid equations for electrons and ions

$$m\frac{\partial}{\partial t}\mathbf{V}_e + m(\mathbf{V}_e\nabla)\mathbf{V}_e = -e\mathbf{E} - \frac{e}{c}\mathbf{V}_e\times\mathbf{B} - \nabla p_e/n + m\nu_{ei}(\mathbf{V}_i - \mathbf{V}_e) \qquad (4.1)$$

$$M\frac{\partial}{\partial t}\mathbf{V}_i + M(\mathbf{V}_i\nabla)\mathbf{V}_i = e\mathbf{E} + \frac{e}{c}\mathbf{V}_i\times\mathbf{B} - \nabla p_i/n - m\nu_{ei}(\mathbf{V}_i - \mathbf{V}_e) \qquad (4.2)$$

Assuming Darwin approximation, introducing conductivity, electron inertia length and expressing electron velocity through current equations could be cast in the following form:

$$\sigma = \frac{ne^2}{m\nu_{ei}}, \quad L_{pe}^2 = \frac{mc^2}{4\pi e^2 n}, \quad \mathbf{J} = \frac{c}{4\pi}\nabla\times\mathbf{B}, \quad \mathbf{V}_e = \mathbf{V}_i - \mathbf{J}/ne \qquad (4.3)$$

$$\frac{\partial}{\partial t}\widetilde{\mathbf{B}} - \nabla\times\left[(\mathbf{V}_i - \mathbf{J}/ne)\times\widetilde{\mathbf{B}}\right] + \nabla\times\frac{c^2}{4\pi\sigma}\nabla\times\mathbf{B} - \frac{c}{ne}(\nabla T_e)\times(\nabla n) = 0 \qquad (4.4)$$

$$\frac{\partial}{\partial t}(M\mathbf{V}_i + m\mathbf{V}_e) + \frac{1}{2}\nabla(m\mathbf{V}_e^2 + M\mathbf{V}_i^2) - \mathbf{V}_i\times\nabla\times(M\mathbf{V}_i + m\mathbf{V}_e) = \frac{1}{nc}\mathbf{J}\times\widetilde{\mathbf{B}} - \frac{\nabla p}{n} \qquad (4.5)$$

Here a new magnetic field function has been introduced

$$\widetilde{\mathbf{B}} = \mathbf{B} + \nabla\times L_{pe}^2\nabla\times\mathbf{B} - \frac{mc}{e}\nabla\times\mathbf{V}_i \qquad (4.6)$$

The ordering of the second and the third term in expression for $\widetilde{\mathbf{B}}$ in units of characteristic scale L is given by $O(L_{pe}^2/L^2)$ and $O(L_{pe}/L\cdot\delta\cdot M_A)$ respectively. Here $M_A = \sqrt{4\pi n M V^2/B^2}$ is characteristic Mach number, $\delta = \sqrt{m/M}$ is small parameter. At usual MHD scales both these terms are obviously small and could be ignored. If one would like to resolve fine scales as well, one can see that the second term becomes comparable to the main one at scale $L_{pe}$, while the third term at much smaller scale $L_{pe}\cdot\delta\cdot M_A$. For many problems Mach number isn't extremely large. We note that for magnetosphere problem characteristic $M_A$ (in contrast to Solar Wind $M_A$) calculated for characteristic field at magnetopause is of order of unity. Next, one can see that the terms $m\mathbf{V}_e^2$ and $m\mathbf{V}_e$ (in comparison to $M\mathbf{V}_i^2$ and $M\mathbf{V}_i$) have ordering $O(L_{pe}/L\cdot 1/M_A)$ and $O(L_{pe}/L\cdot\delta/M_A)$ respectively. Thus, for the purpose of including in the problem fine scales related to electron inertia, it is valid in the first approximation to omit terms of the order $O(L_{pe}/L\cdot\delta)$. Final Hall MHD equations with electron mass taken into account follow as:

$$\frac{\partial}{\partial t}\widetilde{\mathbf{B}} - \nabla\times\left[(\mathbf{V} - \mathbf{J}/ne)\times\widetilde{\mathbf{B}}\right] + \nabla\times\frac{c^2}{4\pi\sigma}\nabla\times\mathbf{B} - \frac{c}{ne}(\nabla T_e)\times(\nabla n) = 0 \qquad (4.7)$$

$$\widetilde{\mathbf{B}} = \mathbf{B} + \nabla\times L_{pe}^2\nabla\times\mathbf{B} \qquad (4.8)$$

$$\frac{\partial n}{\partial t} + \nabla\cdot n\mathbf{V} = 0 \qquad (4.9)$$

$$\frac{\partial p}{\partial t} + (\mathbf{V}\nabla)p + \gamma p\nabla\cdot\mathbf{V} = (\gamma - 1)\cdot\left[\left(\xi + \frac{\eta}{3}\right)(\nabla\cdot\mathbf{V})^2 + \eta(\nabla\times\mathbf{V})^2 + \frac{c^2}{4\pi\sigma}(\nabla\times\mathbf{B})^2\right] \qquad (4.10)$$

$$nM\frac{\partial}{\partial t}\mathbf{V} + nM(\mathbf{V}\nabla)\mathbf{V} = \frac{1}{c}\mathbf{J}\times\widetilde{\mathbf{B}} - \nabla p - \frac{4\pi n}{c^2}\cdot\nabla\left(\frac{1}{2}L_{pe}^2 J^2/n\right) + \eta\nabla^2\mathbf{V} + \left(\xi + \frac{\eta}{3}\right)\nabla\cdot\nabla\cdot\mathbf{V} \quad (4.11)$$

Plasma quasi-neutrality is automatically satisfied because $\nabla\cdot\mathbf{J} = 0$. Viscosity is added for the purpose of resolving shock fronts. Note that the Hall physics is described by a single term $\nabla\times[\mathbf{J}/ne\times\widetilde{\mathbf{B}}]$, while electron mass effects by the effective field $\widetilde{\mathbf{B}}$ and one additional term in momentum equation $\sim \nabla L_{pe}^2 J^2$ which is a consequence of retaining the term $\nabla mV_e^2$. It was checked by comparison that this term doesn't make any significant contribution to results of simulation. However, as long as we include terms of the order $O(L_{pe}/L)$ it should be also kept. To close the set one needs equation for electron temperature. However, further on we ignore the last term in (4.7) assuming that $T_e$ is small and that noncollinearity of temperature and density gradients is small.

The reason of taking into account electron mass effects into Hall MHD is following. The Hall term is notoriously known to be highly unstable in numerical schemes. Because of dispersive nature of whistlers $\omega \sim V_A L_{pi} k^2$ the Courant condition for corresponding waves is $\Delta t < (\Delta r/V_A)\cdot(\Delta r/L_{pi})$. Thus, the required time step for problems with the overall size smaller than $L_{pi}$ is forbiddingly small. However, at small wavelengths $L_{pe}k \sim 1$ dispersion relation changes to $\omega \sim V_A L_{pi} k^2/\sqrt{1+L_{pe}^2 k^2}$, maximum velocity of whistlers is restricted by electron Alfven speed and Courant condition becomes less restrictive $\Delta t < (\Delta r/V_A)\cdot\delta$. It means that if electron mass is consistently taken into account, numerical processing of equations requires time step maximum 45 times smaller than usual MHD one. In magnetosphere problem there is also a factor of substantial density decrease inside cavity. For typical magnetospheric density being order of magnitude smaller than SW density and mesh size $\Delta r = L_{pi}/100$, time saving factor for the scheme with electron inertia included is about 10. Other consideration is that electron inertia smoothers small scale whistler oscillations in a physical way in contrast to artificial means, such as introducing into the numerical scheme relatively large super-viscosity to compensate instability. The mechanism of smoothing is clearly seen in the expression for $\widetilde{\mathbf{B}}$.

To obtain dimensionless set for the problem under consideration we take as typical values downstream plasma velocity, density, kinetic pressure and magnetic field corresponding to this pressure:

$$p_o = n_o M V_o^2,\ B_o^2 = 4\pi p_o,\ \mathbf{B} = \mathbf{B}/B_o,\ \mathbf{V} = \mathbf{V}/V_o,\ n = n/n_o,\ p = p/p_o \quad (4.12)$$

Note that in chosen units of $B_o$ kinetic scales $L_{pi}$ and $R_{Li}$ are exactly equal. Next we define the size of the problem in terms of MHD as the stand off distance where plasma is stopped by magnetic dipole with moment $\mu$:

$$R_M = \left(\mu^2/2\pi p_o\right)^{1/2N_D},\ \mathbf{r} = \mathbf{r}/R_M \quad (4.13)$$

Here $N_D$ is number of dimensions; $N_D = 2$ for the two dimension dipole which is taken in numerical simulation. After defining characteristic length there appears Hall parameter, magnetic and viscose Reynolds numbers:

$$D = R_M/L_{pi},\ S_m = 4\pi\sigma V_o R_M/c^2,\ S_\eta = M V_o n_o R_M/\eta \quad (4.14)$$

Dimensionless set follows as:

$$\frac{\partial}{\partial t}\tilde{\mathbf{B}} - \nabla \times \left[\left(\mathbf{V} - \frac{1}{D}\mathbf{J}/n\right) \times \tilde{\mathbf{B}}\right] + \frac{1}{S_m}\nabla \times \nabla \times \mathbf{B} = 0 \tag{4.15}$$

$$\tilde{\mathbf{B}} = \mathbf{B} + \delta^2 \cdot \nabla \times \frac{1}{n}\nabla \times \mathbf{B}, \quad \mathbf{J} = \nabla \times \mathbf{B} \tag{4.16}$$

$$\frac{\partial n}{\partial t} + \nabla \cdot n\mathbf{V} = 0 \tag{4.17}$$

$$\frac{\partial p}{\partial t} + (\mathbf{V}\nabla)p + \gamma p \nabla \cdot \mathbf{V} = (\gamma - 1) \cdot \left[\frac{1}{S_\eta}(\nabla \cdot \mathbf{V})^2 + \frac{1}{S_\eta}(\nabla \times \mathbf{V})^2 + \frac{1}{S_m}(\nabla \times \mathbf{B})^2\right] \tag{4.18}$$

$$n\frac{\partial}{\partial t}\mathbf{V} + n(\mathbf{V}\nabla)\mathbf{V} = \mathbf{J} \times \tilde{\mathbf{B}} - \nabla p - n \cdot \delta^2 \cdot \nabla\left(\frac{1}{2}\mathbf{J}^2/n^2\right) + \frac{1}{S_\eta}\left(\nabla^2\mathbf{V} + \nabla \cdot \nabla \cdot \mathbf{V}\right) \tag{4.19}$$

Dimensionless magnetic moment is given by $\mu = \mu/B_o \cdot (R_M)^{-ND} = 1/\sqrt{2}$. The small parameter $\delta^2 = 1/1836$ is electron to ion mass ratio. To save the space we combined bulk and dynamic viscosities by taking $\xi = 2\eta/3$. To resolve shock wave structure that satisfies Rankine-Hugoniot conditions it is necessary to take simulation viscose number $S_\eta$ to be not much larger than inverse grid size. Magnetic Reynolds number in simulations was sufficiently large $S_m \geq 400$. Except shock and magnetopause, viscosity and conductivity doesn't play significant role in the processes of interest. At the input boundary conditions of the SW are imposed $n = 1$, $V_x = -1$, $p = M_s^{-2}/\gamma$ where $M_s = V_o/C_s$, $C_s = \sqrt{\gamma T_o/M}$. In calculations we take $\gamma = 5/3$, $M_s = 7$. For most calculations no magnetic field in the external plasma flow is considered $B_{IFM} = 0$. Initially simulation box is filled with stationary rarified and cold plasma $n \leq 0.1$. Practically steady state of interaction with SW is reached after several tens to several hundreds of characteristic times. For time integrating of density and pressure a direct implicit "upwind" discretization in conservative form was used while for velocity and magnetic field - implicit Lax-Wendorff scheme was adopted. The latter greatly increases numerical stability in the vicinity of dipole origin where Alfven speed is very large. In the region of about several mesh points around dipole center density and pressure were kept constant and small to limit accumulation of plasma reaching the dipole through cusps.

For the 2D problem it is convenient to employ component of vector potential and magnetic field along the translational coordinate y:

$$\mathbf{B}_\perp = -\mathbf{e}_y \times \nabla A_y, \quad \mathbf{J}_\perp = -\mathbf{e}_y \times \nabla B_y, \quad J_y = -\nabla \cdot \nabla A_y \tag{4.20}$$

Effective values that take into account electron mass are given by

$$\tilde{A}_y = A_y - \frac{\delta^2}{n}\nabla \cdot \nabla A_y, \quad \tilde{B}_y = B_y - \delta^2 \nabla \cdot \frac{1}{n}\nabla B_y \tag{4.21}$$

In these terms equation (4.15) transforms to:

$$\frac{\partial}{\partial t}\tilde{B}_y + \nabla \cdot \left(\mathbf{V}_\perp \tilde{B}_y\right) + \frac{1}{S_m}\nabla^2 B_y = \left(\tilde{\mathbf{B}}_\perp \cdot \nabla\right) \cdot \left(V_y - J_y/nD\right) \tag{4.22}$$

$$\frac{\partial}{\partial t}\tilde{A}_y + (\mathbf{V}_\perp \nabla)\tilde{A}_y + \frac{1}{S_m}\nabla^2 A_y = -\frac{1}{nD}\left(\tilde{\mathbf{B}}_\perp \cdot \nabla\right)B_y \tag{4.23}$$

## 5. Results of numerical simulation

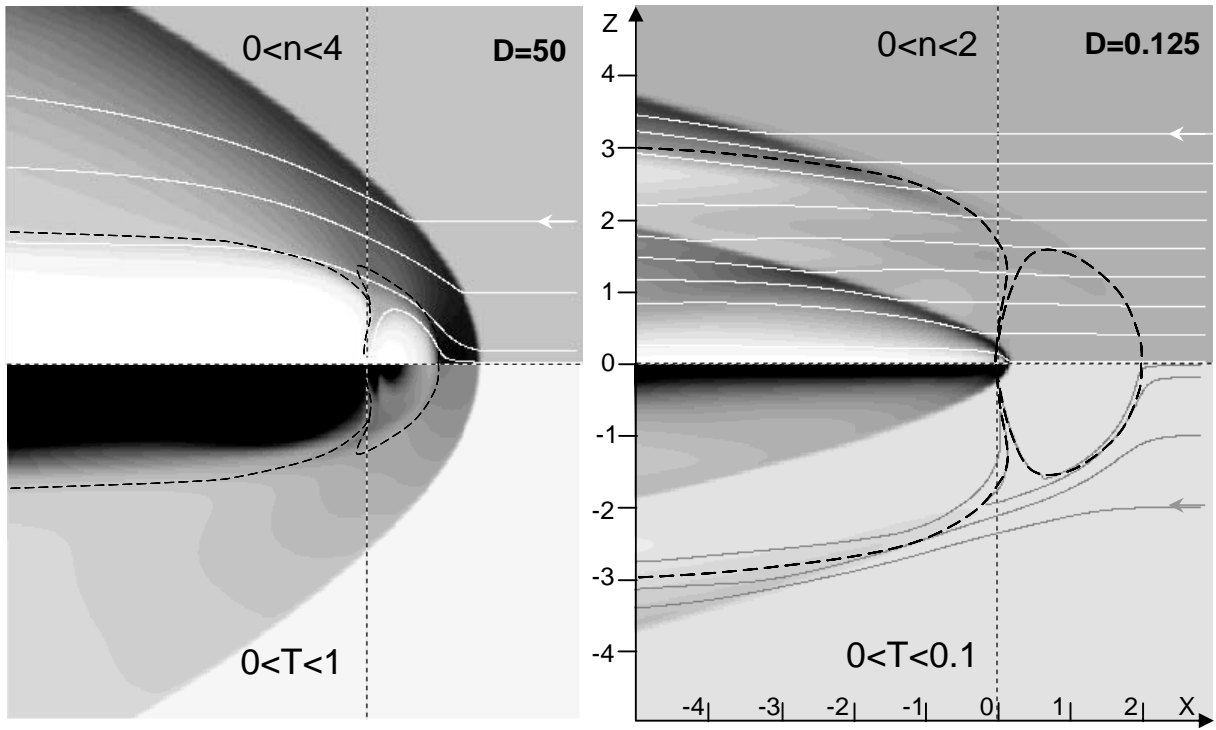

**Figure 8** General structure of magnetosphere in MHD regime (left) and Hall regime (right). Grayscale plots of plasma density (upper half of pictures) and temperature (bottom half). Ranges indicate maximum white and maximum black colors. Also shown ion streamlines (solid white lines), electron streamlines (solid grey lines) and magnetopause boundary (dashed black line).

In figure 8 plots of plasma density and temperature are shown. As these values are symmetric in respect to X axis they are combined in one picture. At sufficiently large D Hall effects are negligible and the left plot presents MHD regime of interaction. One can see a clear bow shock and cavity of rarified and hot plasma around dipole and in the tail. SW plasma doesn't penetrate magnetopause boundary. Totally different picture is observed in strongly kinetic regime. Magnetopause boundary is positioned significantly farther from the dipole and there is no preceding bow shock. Plasma deceleration is very small at the sub-solar point and becomes more visible in the tail. Strong ion deflection is observed only in a close vicinity of the dipole. This deflection is related to strong plasma perturbation generated close to the dipole and extending far in the tail. Electrons don't penetrate inside magnetosphere and overflow dipole around the magnetopause boundary. We note that in previous case ion and electron streamlines are effectively equal.

Comparative profiles of magnetosphere for large, order of unity and small Hall parameter is shown in figure 9. In MHD regime a bow shock, magnetopause and magnetosheath in between where thermal pressure is close in value to kinetic pressure of SW are clearly seen. Magnetopause position is very close to the expected stand off sub-solar distance. In Hall regime D=0.125 magnetopause is found at a distance twice larger. The jump of field is correspondingly small $\Delta B_z \approx \sqrt{2}/R_m^2$. Plasma deceleration at magnetopause is negligible and pressure varies adiabatically: $nV_x = \text{const}$, $p \sim n^\gamma$, $\Delta V_x \approx \Delta B_z^2/2$. Thus, no shock develops upstream. However, shocked region appears close to the dipole origin where plasma eventually stops. The nature of this region will be discussed later. Intermediate regimes clearly demonstrate how these features develop. At D=1 the width of magnetosheath contracts and plasma significantly penetrates beyond magnetopause, while at D=0.5 bow shock disappears altogether.

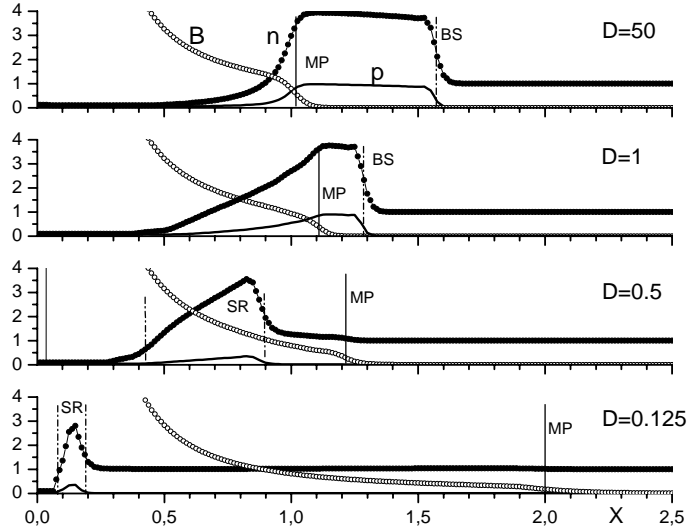

**Figure 9** Profiles of magnetic field (o), plasma density (●) and pressure (solid line) along X axis for four values of parameter D. Thin vertical lines indicate bow shock (BS), sub-solar magnetopause (MP) and a region of shocked plasma (SR) which develops inside magnetosphere in kinetic regimes.

We attribute all these differences to the Hall out of plane field $B_y$. Its spatial structure for large, intermediate and small parameter D is shown in the next figure 10. Maximum value of $B_y$ in the first case is significantly smaller than in others. Another essential difference is in spatial structure. In MHD case Hall field has fine pattern following magnetopause current. It shows positive and adjacent negative layers in both hemispheres. In kinetic regime Hall field smoothly fills whole magnetosphere and is everywhere positive in the North and negative in the South hemisphere. Detailed analysis reveals that it is generated by magnetopause current and convected by plasma inside magnetosphere. In the intermediate regime Hall field closely follows magnetopause and is dominantly positive in the North and negative in the South hemisphere. Middle picture shows striking similarity to experimental distribution measured at the same value of parameter D (figure 1C). They are also in good quantitative agreement – maximum values in units of characteristic field $\sqrt{4\pi p_o}$ (4.12) are 0.25 in experiment and 0.22 in simulation.

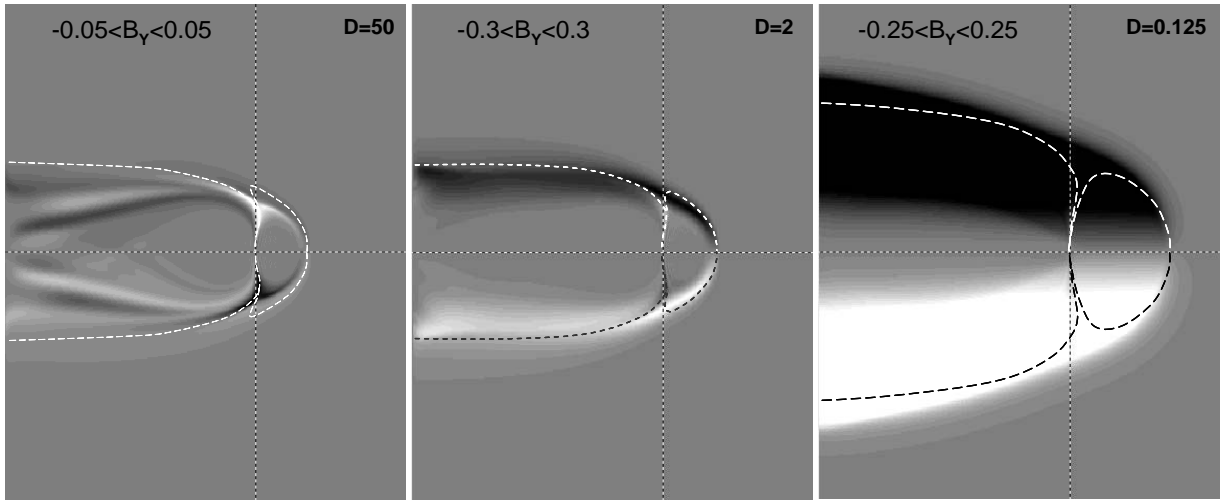

**Figure 10** Grayscale plots of $B_Y$ component of magnetic field for MHD (left), intermediate (middle) and Hall regime (right). White color corresponds to negative and black to positive values. Dashed line shows magnetopause. Spatial dimensions are the same as in figure 8.

Why the Hall field so much affects magnetosphere in kinetic regime is illustrated in figure 11 where profiles of plasma and current velocity are shown. One can see that at magnetopause positioned at $X \approx 2$ plasma slightly decelerates and electric current related to $B_y$ field sharply jumps to compensate ion current. It appears that inside all of magnetosphere, in the frontal part, in the tail and high-latitude regions except cusps, in plane ion velocity is equal to current velocity $\mathbf{V} = \mathbf{J}/ne$ while electron velocity is close to zero. Because of this magnetic field isn't advected, even if ions freely move across magnetosphere. In figure 11 there is also shown the out of plane component of plasma velocity which is another manifestation of the Hall effect. It is generated by magnetic force $\sim J_x B_z/n \approx V_x B_z$, reaches value comparable to upstream velocity and corresponds by sign to ion gyrorotation.

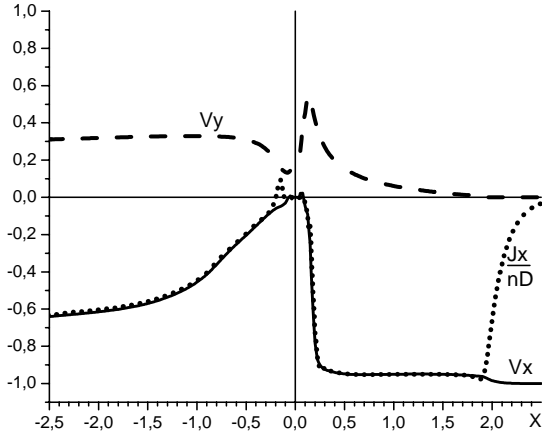

**Figure 11** Profiles of plasma velocity (solid) and current velocity $J_x/nD$ (dotted) along X axis for parameter D=0.125. Dashed line shows out of plane velocity $V_Y$.

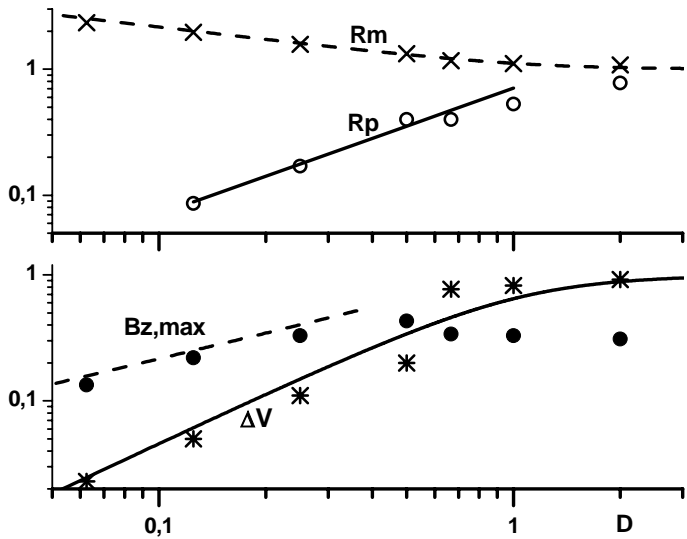

**Figure 12**
Sub-solar magnetopause position (X), closest plasma approach to the dipole origin (O), deceleration of plasma across magnetopause (✳) and maximal out of plane magnetic field (●) in dependence on the parameter D. Solid and dashed lines are given by analytical expressions (5.1, 5.2).

In figure 12 general characteristics of mini-magnetosphere in dependence on the Hall parameter are shown. They are compared with analytical estimations introduced in section 3. In normalized units and for 2D scaling of magnetopause position the values of interest are given by:

$$\Delta v^{3/2} = 2D^2 v_p \big/ (1 + v_p), \quad R_m = \Delta v^{-1/4}, \quad B_{z,\max} = D \cdot R_m \tag{5.1}$$

Parametric study reveals that at small D closest plasma approach is approximately equal to Stoermer radius which is the closest approach of test ions impinging perpendicular on 2D dipole. In normalized units it is given by

$$R_{St}/R_M = e\mu/(V_0 McR_M) = D/\sqrt{2} \qquad (5.2)$$

One can see that at sufficiently small D analytical estimations are in a good agreement with numerical simulation. In the intermediate range $D \approx 1$ shocked plasma and Hall effects are strongly intermingled. Bow shock disappears between D=0.7 and 0.5, exactly in the range where penetration velocity increases above $1/4$. At $D = 2$ Hall effects are relatively unimportant though out of plane field is still sufficiently large.

Next we address the question of plasma pile up and heating in a small region of order of Stoermer radius near the dipole. In (Fujita 2004) it was found out that effective cross section of plasma strong deflection scales as square of $R_{St}$ in the limit of small D. However, no strong disturbances on this scale are reported in PIC simulations. To access the limitations of Hall MHD and PIC approaches for this particular problem we employ a model of test particles. As has been shown above, in strongly kinetic regime ions freely penetrate magnetosphere and move there as particles orbiting in magnetic field. Introducing a uniform flux of ions impinging from infinity on the magnetic dipole $f \sim \exp\left[-M(\mathbf{V}-\mathbf{V}_0)^2/2T_0\right]$, $\mathbf{V}_0 = -\mathbf{e_x}V_0$ and calculating individual trajectories we can find density and temperature as corresponding moments of distribution function $n = \int f(\mathbf{V})$, $3nT = \int (\mathbf{V}-\langle\mathbf{V}\rangle)^2 f$. Result obtained by use of $3 \cdot 10^7$ particles with initial temperature $T_0 = 0.012 \cdot MV_0^2$ is shown in figure 13. Strong plasma pile up $n_{max} \approx 2.2$, heating $T_{max} \approx 0.1$ and sufficiently large thermal pressure $p_{max} \approx 0.2$ do develop at the Stoermer limit boundary. Density perturbation is distributed in a thin layer, while effective heating and thermal pressure (in a sense of corresponding moment of distribution function) spreads over wide region around dipole. Note that perturbation extends in the tail as well. It might be concluded that Hall MHD isn't essentially wrong and at least qualitatively captures the features of the shocked region. However, how exactly reflected bunches of ions will interact with main flow and thermalize, on what scales and through which instabilities could be answered only by PIC analysis.

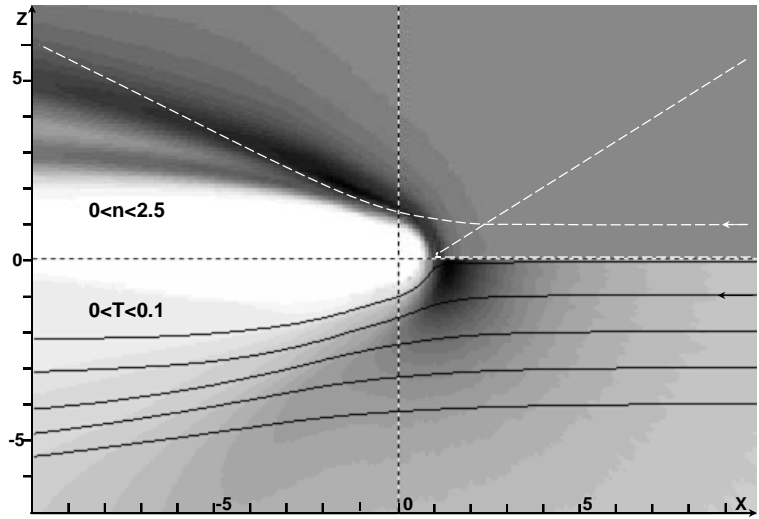

**Figure 13** Grayscale plots of plasma density (upper half of picture) and temperature (bottom half) calculated by test particles model. Also shown mean velocity streamlines (solid black lines) and a couple of ion trajectories (dashed white). Spatial scale is in units of Stoermer radius.

Important question is how the IMF influences the above features. Results of simulation in strongly kinetic regime and in presence of SW magnetic field $B_{IMF,x} = 0.1$, $B_{IMF,z} = 0.1$ corresponding to Mach number $M_A = 7$ are shown in figure 14. Black vectors show bifurcated magnetic field lines which divide closed magnetosphere from open lines and unbroken SW lines.

Density plot and ion streamlines (left picture) are similar to figure 8 (right) obtained without IMF. Namely, ions freely penetrate magnetosphere, there is no upstream bow shock and strong density perturbation is generated in a close vicinity of the dipole where ions are deflected.

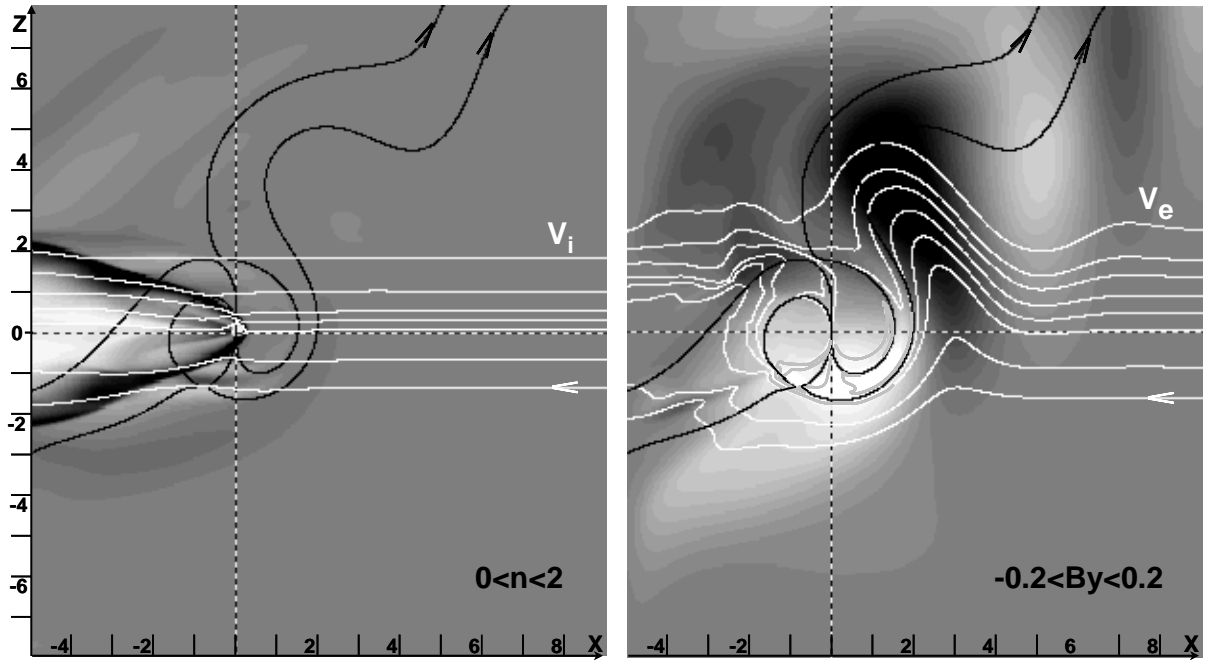

**Figure 14** Mini-magnetosphere in presence of IMF inclined 45° Northward and Sunward. Hall parameter D=0.125. Grayscale plots of density (left) and out of plane magnetic field (right). Black lines show magnetic field lines, white lines – ion and electron streamlines.

However, Hall field and electron streamlines reveal a much more complex picture. The structure of $B_y$ component, though generally positive above and negative below equator, is strongly distorted by inclination along the IMF and by spatial oscillations. Electron streamlines also experience oscillations. They either skirt around magnetosphere or pass close to the reconnection sites. In case of Northward $B_{IMF,z}$ the reconnection process with dipole field develops over poles. As it is well known, at scales below ion inertia length reconnection is supported by electrons while ion dynamics is inessential. In a close up view of merging sites it was verified that electrons flow towards X-point driving magnetic lines from each side and outflow in opposite directions away from the X-point approximately along field lines. On the other hand, one can see that ion streamlines pass across merging sites undisturbed. One of the findings of Hall mediated reconnection is that merging rate doesn't depend on the actual dissipation mechanism as long as this dissipation is small. Thus, sufficiently large magnetic Reynolds number used in simulations couldn't affect merging rate and general structure of magnetosphere. Despite of the reconnection process and distortion of Hall field, one can see that most region of inner magnetosphere is shielded against direct penetration of SW electrons. It may be concluded that on the scale of pressure balance distance IMF doesn't change the properties of mini-magnetosphere described above.

Finally, we consider the tail of mini-magnetosphere. It extends for a long distances in terms of either pressure balance distance or ion inertia length as is shown in figure 14. Without IMF magnetic field lines of dipole form a stretched cavity. The lobe field rather slowly falls off along the tail. Out of plane field also smoothly extends along the tail. The noticeable feature is that it is several times larger than the lobe field. Density and pressure perturbations show two wakes. The inner wake inside magnetosphere is generated in a close vicinity of the dipole discussed in relation to figure 13. The second outside wake is a shock generated behind the dipole. Its development and ions deflection at the front are clearly seen in close up view in figure 8. Shock appears behind the dipole because in the tail magnetopause current changes much slower along magnetic field lines than at the front and, according to (4.22-23), Hall effects are small. To see weather the outside perturbation is a shock we plot in figure 16 profiles across the tail. For comparison the MHD case is presented as well. MHD regime shows a clear bow shock at which pressure and ideal gas entropy $S = \ln(p/n^\gamma)/(\gamma-1)$ jump. Magnetosheath region is bounded by a cavity at which tail field exists and where plasma is rarified and hot. Magnetopause is identified as a boundary of field increase. At this boundary variations of magnetic and thermal pressures nearly exactly compensate each other $B_x^2/2 + p \approx \text{const}$.

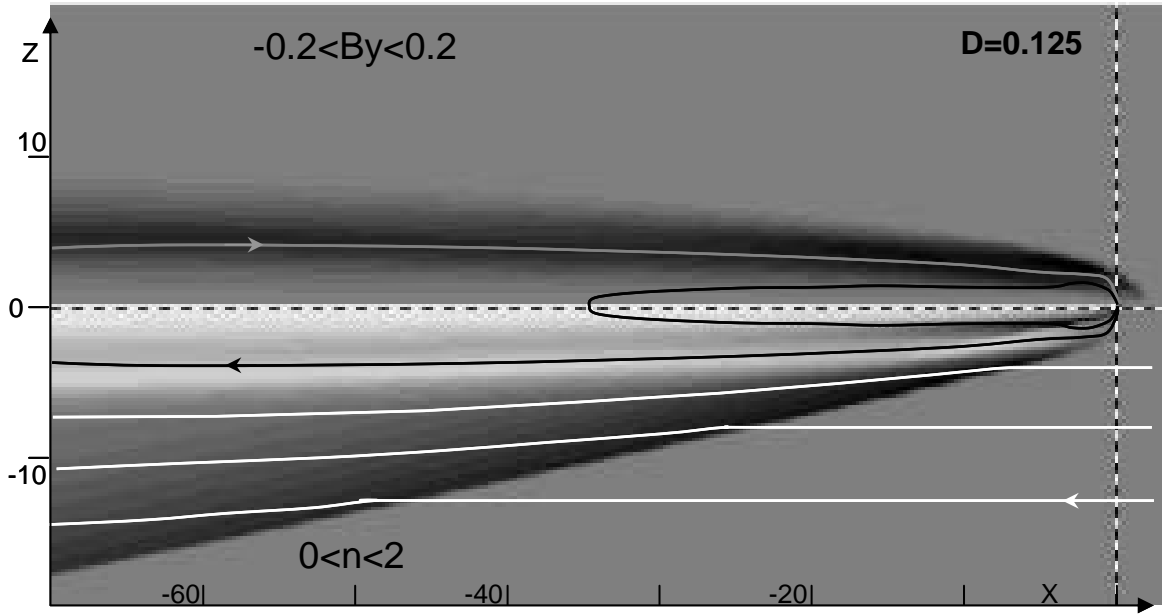

**Figure 15** Tail of magnetosphere in the Hall regime D=0.125. Grayscale plots of $B_Y$ component (upper half) and density (bottom half). Black lines show magnetic field lines, white lines – ion streamlines.

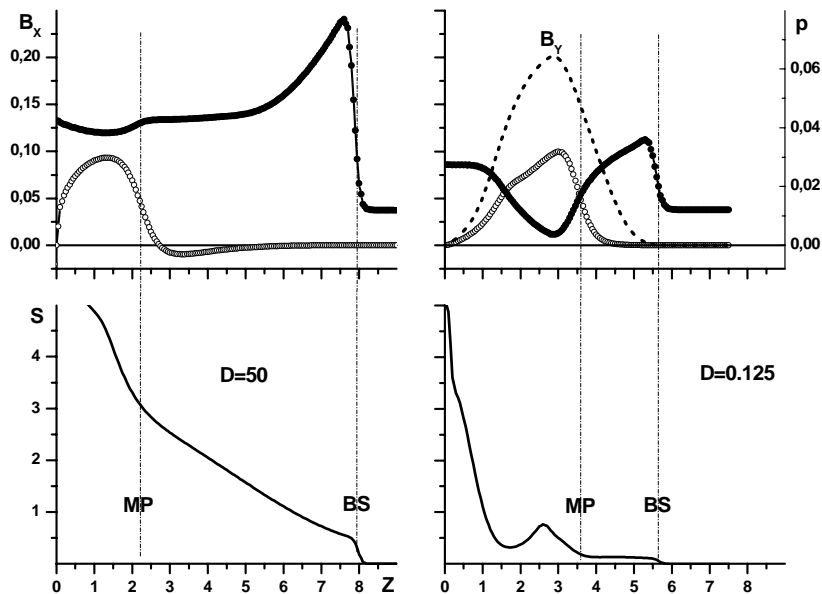

**Figure 16** Cross-tail profiles at position X=-14 for MHD (left panels) and Hall regime (right panels). Upper panels – $B_X$ component of magnetic field (o) and plasma pressure (●). Lower panels – enropy (solid lines). For the Hall regime out of plane $B_Y$ component is also shown by dashed line. Thin vertical lines indicate bow shock (BS) and magnetopause (MP).

Hall regime shows the same characteristics: at the leading front there is entropy jump, albeit significantly smaller. At magnetopause boundary of dipole field the thermal pressure falls off. However, in this case thermal pressure is balanced by total field $(B_x^2 + B_y^2)/2 + p \approx const$. Thus, mostly the out of plane field rather than dipole field serves as the obstacle to SW plasma. $B_y$ component (right upper panel of figure 16) fills lobe and in part magnetosheath region. Note also that exactly like it was observed at the frontal magnetosphere, magnetopause position is significantly farther off and magnetosheath is thinner, and for smaller values of Hall parameter such tendencies are more pronounced. In summary, tail of mini-magnetosphere exhibits features of weak magnetosonic shock on the one hand, and almost 90° out of plane inclination of magnetic field that can be interpreted as a whistler wake on the other.

In presence of IMF the large scale structure of mini-magnetosphere dramatically changes. Large scale picture corresponding to figure 14 is shown in figure 17. While at the quasi-perpendicular front of perturbation cone the upstream region is undisturbed, at the quasi-parallel front there are significant oscillations of magnetic field that run away far upstream. This is explained by the fact that perpendicular to IMF wave velocity is magnetosonic speed which is smaller than upstream speed. On the other hand, parallel to IMF whistler wave velocity $V_A L_{pi} k \approx 0.7$ (for the observed wavelength $\lambda \approx 10$ along the field line passing close to the dipole) is equal to upstream velocity along IMF and a standing wave pattern is formed.

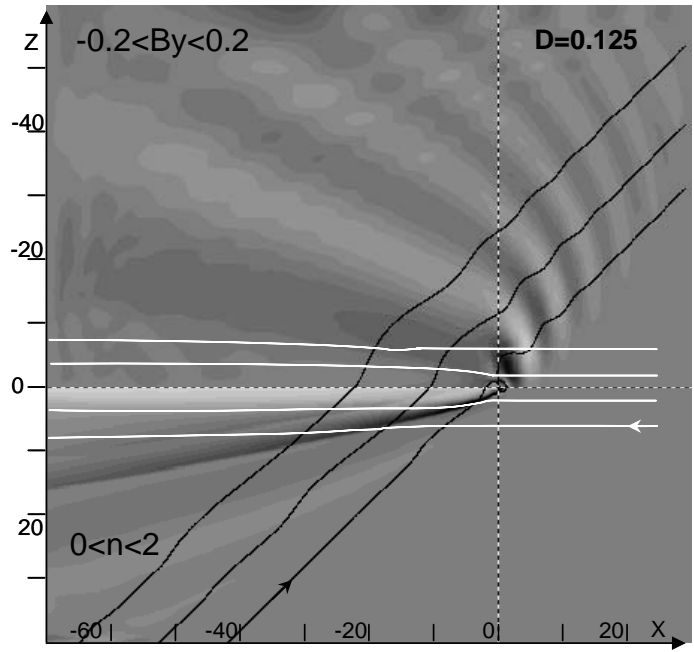

**Figure 17** Large scale picture of mini-magnetosphere in presence of IMF inclined 45° Northward and Sunward. Hall parameter D=0.125. Grayscale plots of out of plane magnetic field (upper half) and density (bottom half). Black lines show magnetic field lines, white lines – ion streamlines.

The other noticeable feature is that in the tail the Hall field disintegrates into oscillations and doesn't have simple dual structure as in case without IMF. The same is true for variations of other field components. To understand the nature of density and pressure wake in figure 17 we plot in figure 18 cross-tail profiles for cases of MHD and Hall regimes with the same IMF. In the MHD regime out of plane current clearly indicates positions of bow shock and magnetopause. Inside the bow shock cone pressure force is balanced by magnetic force. In the Hall regime current $J_y$ shows only oscillations and by value is order of magnitude smaller. Thus, magnetic force is inessential and it was checked that pressure is balanced by inertia term not only at the cone front but inside it as well. The entropy jump at the front (not shown) is order of magnitude smaller. It proves that density and pressure wake in this case is a magnetosonic perturbation, not a shock. If IMF lines cross the tail the Hall effects are always large at $D \ll 1$ and magnetosonic waves are strongly coupled to whistlers. Because whistlers are much faster any large perturbation of pressure and density disintegrates into whistler waves.

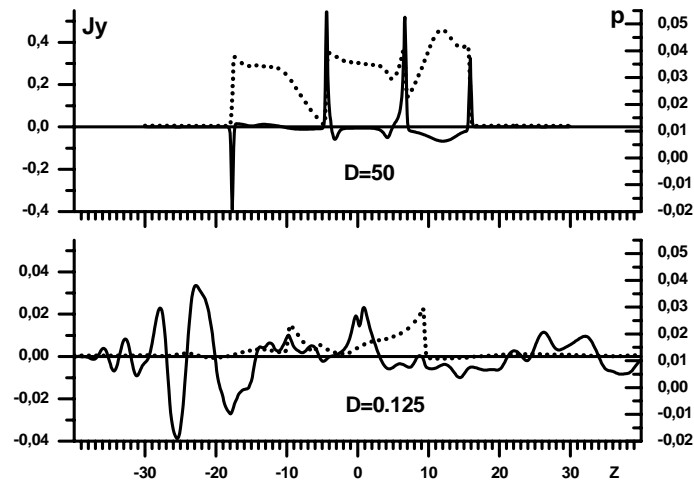

**Figure 18** Cross-tail profiles at position X=-35 for MHD (upper panel) and Hall regime (bottom panel). Solid curves – out of plane current, dotted curves – pressure.

**6. Discussion and conclusions**

Three approaches have been used in the present work to study properties and physics of mini-magnetosphere. Conducted laboratory experiments are a most close representation but due to restricted interaction time they lack bow shock. Numerical model includes two-fluid physics and reproduces bow shock but is two dimensional and disregards kinetic effects. Analytical model describes a process which is behind the observed features but is necessarily very simplified. Despite of the differences all approaches revealed essentially the same picture and supplemented each other in details. Namely, when ion inertia length is larger than pressure balance distance magnetopause shifts farther away from the dipole, jump of field lessens and plasma penetrates into magnetosphere to be stopped at Stoermer limit. Out of plane component of magnetic field directed along magnetopause current is found to be behind such dramatic change. Experimentally observed spatial structure and independence on the sign of magnetic moment give direct evidence that this field is generated by magnetopause current via Hall term. Quantitative analytical estimates of sub-solar magnetopause position, penetration velocity and Hall field are consistent with results of numerical simulation and experimental data.

Developed model explains why a mini-magnetosphere is so much different. At magnetopause boundary the Chapman-Ferraro current generates magnetic field along its direction as described by the Hall term in the Ohm' law. The resulting new current system advects magnetic field in meridian plane, as described by the same Hall term. In steady state to cancel this additional advection plasma velocity tends to be equal to current velocity which in effect means two things. First, plasma penetrates into magnetosphere, and because the jump of kinetic pressure lessens the magnetopause position correspondingly shifts away from dipole. Second, plasma dynamics inside magnetosphere is described by a particle motion law in the dipole field. In other words, Hall currents tend to cancel electric fields so ions move only under magnetic force. In this case plasma is stopped at Stoermer particle limit. Disappearance of bow shock is explained by penetration of plasma across magnetopause. With increase of Hall currents penetration velocity also increases and, when it exceeds maximum possible velocity in magnetosheath region as determined by Rankine-Hugoniot relations, a standing shock cannot exist.

Disappearance of bow shock and penetration of plasma beyond the pressure balance distance was observed in a number of PIC simulations sited above. In (Fujita 2004) it was also deduced that in kinetic regime plasma is deflected at Stoermer limit. Presence of global out of plane magnetic field generally positive in the North and negative in the South hemisphere can be found in (Blanco-Cano *et al* 2006) though it isn't discussed in any detail. However, there is one important feature not described in previous PIC simulations. While bow shock disappears in strongly kinetic regime, there remains magnetopause as a boundary of dipole field. SW electrons overflow magnetosphere around this boundary and don't directly penetrate inside as ions do. This is essentially novel feature of mini-magnetosphere that could be of fundamental and practical interest. We note that there is no contradiction with charge neutrality condition. It is automatically fulfilled in Hall MHD model and electrons that neutralize ions inside magnetosphere constitute quasi-stationary population formed in the course of magnetosphere formation. At changing SW conditions Hall currents adjust to replenish this population by SW electrons as in non-stationary magnetosphere ion current isn't necessarily equal to electric current. In reality, a number of processes can contribute to exchange between SW and magnetospheric electrons such as small scale instabilities and reconnection in presence of IMF. If these indirect processes are slow enough, magnetospheric population of electrons might develop features distinctly different from SW. As demonstrated in figure 13, mini-magnetosphere is filled by ions reflected near Stoermer limit. Such reflected bunches are a source of various instabilities and waves, and heating of electrons is to be expected. All these processes could be studied only by kinetic models. Hall MHD gives a general picture of mini-magnetosphere as a starting point of more elaborate analysis.

It might seem intuitively obvious that ions should penetrate beyond magnetopause by as much as their gyroradius. It might be argued that the width of transition layer simply couldn't be smaller than gyroradius and when it is large enough, the closest boundary should be the Stoermer limit. However, the fine point is that plasma dynamics inside mini-magnetosphere, while could be viewed as a gyromotion, is determined not by kinetic effects as described by distribution function but by two-fluid physics. Manifestation of this physics – generation of Hall magnetic field – isn't obvious at all.

In the numerical simulation we studied also the tail of mini-magnetosphere and the effect of IMF. Without IMF weak magnetosonic shock is observed behind the dipole and the lobe magnetic field is dominated by out of plane Hall component. Thus, a spacecraft crossing the tail will see reversing field directed almost perpendicular to the tail orientation and dipole moment. In the presence of IMF it was found that on the scale of pressure balance distance the main features of mini-magnetosphere remain the same though spatial structure becomes distorted along IMF direction. However, on large scales IMF influence is dramatic. In this case there is no shock in the tail, either quasi-perpendicular or quasi-parallel. Perturbations generated near the dipole propagate in the tail as magnetosonic and whistler waves. Whistler waves, being fast, also propagate far upstream of the dipole. This picture well agrees with detailed analysis based on PIC simulation (Omidi *et al* 2002, Blanco-Cano *et al* 2006). Lunar Prospector magnetometer collected a large dataset of events of nearly monochromatic circular low frequency waves clearly associated with lunar crustal magnetic fields (Halekas *et al* 2006). It was deduced that inferred wave properties are completely consistent with propagating whistlers or phase standing whistler wake generated either at a shock surface or in direct interaction of SW with crustal fields.

So far related observations from space are rather scanty and incomplete. It is yet to be seen that numerical and analytical studies of the subject are tested against natural mini-magnetospheres formed by the Solar Wind around weakly magnetized bodies.

Acknowledgements: This work was supported by SB RAS Research Program grant II.8.1.4, Russian Fund for Basic Research (grants 09-02-13578, 09-02-00492, 09-08-00970) and OFN RAS Research Program 15.

## References

R Bamford, K J Gibson, A J Thornton, J Bradford, R Bingham, L Gargate, L O Silva, R A Fonseca, M Hapgood, C Norberg, T Todd and R Stamper 2008 The interaction of a flowing plasma with a dipole magnetic field: measurements and modelling of a diamagnetic cavity relevant to spacecraft protection *Plasma Phys. Control. Fusion* **50**(12) 124025

Bernhardt PA, Roussel-Dupre R A, Pongratz M B, Haerendel G, Valenzuela A 1987 Observations and theory of the AMPTE magnetotail Barium releases *J. Geophys. Res*. **92A** 5777-94

Blanco-Cano X, N Omidi and C T Russell 2003 Hybrid simulations of solar wind interaction with magnetized asteroids: Comparison with Galileo observations near Gaspra and Ida *J. Geophys. Res.* **108**(A5) 1216

Blanco-Cano X, N Omidi and C T Russell 2004 Magnetospheres: How to make a magnetosphere *Astronomy & Geophysics* **45** (3) 3.14-3.17

Blanco-Cano X, N Omidi and C T Russell 2006 Macrostructure of collisionless bow shocks: 2. ULF waves in the foreshock and magnetosheath *J. Geophys. Res.* **111** A10205

Cohen LG, SKF Karlsson 1969 Experimental studies of the interaction between collisionless plasmas and electromagnetic fields *AIAA Journal* **7** 1446-53

Fruchman A and Maron Y 1991 Fast magnetic-field penetration into plasmas due to the Hall field *Phys. Fluids* **B3** (N7) 1546-51

Fujita K Particle simulation of moderately-sized magnetic sails 2004 *J. Space Technol. Sci.* **20**(2) 26-31

Gargaté L, R Bingham, R A Fonseca, R Bamford, A Thornton, K Gibson, J Bradford and L O Silva 2008 Hybrid simulations of mini-magnetospheres in the laboratory *Plasma Phys. Control. Fusion* **50**(7) 074017

Halekas J S, D A Brain, D L Mitchell, R P Lin 2006 Whistler waves observed near lunar crustal magnetic sources *J. Geophys. Res.* **33** L22104

Halekas J S, G T Delory, D A Brain, R P Lin, D L Mitchell 2008 Density cavity observed over a strong lunar crustal magnetic anomaly in the solar wind: A mini-magnetosphere? *Planetary and Space Science* **56** 941–6

Hassam A B and J D Huba 1987 Structuring of the AMPTE magnetotail Barium release *Geophys. Res. Lett.* **14**(1) 60-3

Mandt M E, R E Denton and J F Drake 1994 Transition to whistler mediated magnetic reconnection *Geophys. Res. Lett.* **21** 73-6

Mikhailichenko V A, Morozov A I, Smirnov V A and Tilinin G N 1973 Investigation of "Contour" Oscillations in Accelerators with Closed Electron Drift and Extended Acceleration Zone *Proc. 2nd All-Union Conference on Plasma Accelerators and Ion Injectors*, *Minsk, 1973*, p12, In Russian.

Mordovskaya V G, V N Oraevsky, V A Styashkin and J Rustenbach 2001 Experimental Evidence of the Phobos Magnetic Field *JETP Letters* **74** 6 293–7

Nagamine Y and H Nakashima 1999 Analysis of plasma behavior in a magnetic thrust chamber of Laser Fusion Rocket *Fus. Techn.* **35** 62-70

Nikitin S A and Ponomarenko A G 1995 Energetic criteria of artificial magnetosphere formation *Journal of Applied Mechanics and Technical Physics* **36** N4 483-7

Omidi N, X Blanco-Cano, C T Russell, H Karimabadi and M Acuna 2002 Hybrid simulations of solar wind interaction with magnetized asteroids: General characteristics *J. Geophys. Res.* **107**(A12) 1487, doi:10.1029/2002JA009441

Okada S, K Sato and T Sekiguchi 1981 Behaviour of laser-produced plasma in a uniform magnetic field – plasma instabilities *Jpn. J. Appl. Phys.* **20** 157-65

Ponomarenko A G, Zakharov Yu P, Antonov V M, Boyarintsev E L, Melekhov A V, Posukh V G, Shaikhislamov I F and Vchivkov K V 2008 Simulation of strong magnetospheric disturbances in laser-produced plasma experiments *Plasma Phys. Control. Fusion* **50** 074015

Shabansky V P, Veselovsky I S, Koval A D, Ryabushka S B, Us A A, Shuvalov V A 1989 Application of artificial magnetic field in space research *Proc 18 Gagarin Scientific Talks in Cosmonautics and Aviation* Moscow 207-8 (in Russian)

Spillantini P, F Taccetti, P Papini, L Rossi 2000 Radiation shielding of spacecraft in manned interplanetary flights *Nuclear Instr. and Methods in Phys. Res. A* **443** 254-63

Winglee R M, J Slough, T Ziemba, A Goodson 2000 Mini-Magnetospheric Plasma Propulsion - Tapping the energy of the solar wind for spacecraft propulsion *J. Geophys. Res.* **105**(A9) 21067-77

Saito Y et al 2010 In-flight Performance and Initial Results of Plasma Energy Angle and Composition Experiment (PACE) on SELENE (Kaguya) *Space Sci. Rev.* **154**(N1-4) 265-303

Zakharov Yu P, A M Orishich, A G Ponomarenko and V G Posukh 1986 Effectiveness of the slowing of expanding clouds of diamagnetic plasma by a magnetic field (experimental) *Fiz. Plazmy* **12** 1170

Zakharov Yu P, V M Antonov, E L Boyarintsev, K V Vchivkov, A V Melekhov, V G Posukh, I F Shaikhislamov, A G Ponomarenko, I S Veselovsky, A T Lukashenko 2009 On the Interaction Effects of Ionospheric Plasma with Dipole Magnetic Field of the Spectrometer AMS-02 Moving Onboard of International Space Station *Proc. Int. Conf. on "Space Science & Communication (26-27 October 2009, Port Dickson, Malaysia)* 96-101